\documentclass[a4paper,10pt]{article}
\usepackage{enumerate}
\usepackage{color}
\usepackage[utf8]{inputenc} 
\usepackage[english]{babel}
\usepackage[T1]{fontenc}
\usepackage{graphicx}
\usepackage{amsfonts,amssymb,amsmath,latexsym,amsthm}
\usepackage{textcomp}
\usepackage[pdftex]{hyperref}
\usepackage{geometry}
\title{Wave evolution within the Cubic Vortical Whitham equation}
\author{Marcelo V. Flamarion$^{1}$ and Efim Pelinovsky$^{2,3}$}
\date{}

\begin{document}
\maketitle
\begin{center}
{\footnotesize $^1$ Departamento Ciencias—Secci{\' o}n Matem{\' a}ticas, Pontificia Universidad Cat{\' o}lica del Per{\' u}, Av. Universitaria 1801, San Miguel 15088, Lima, Peru  \\
mvellosoflamarionvasconcellos@pucp.edu.pe}

\vspace{0.3cm}
{\footnotesize $^{2}$Faculty of Informatics, Mathematics and Computer Science, HSE University, Nizhny Novgorod 603155, Russia.

$^{3}$ Gaponov-Grekhov Institute of Applied Physics, Nizhny Novgorod, 603122, Russia.}


\end{center}


\begin{abstract}
In this work, we study the evolution of disturbances within the framework of the Cubic Vortical Whitham (CV-Whitham) equation, considering both positive and negative cubic nonlinearities. This equation plays important role for description of the wave processes in the presence of shear flows. We find well-formed breather-type structures arising from the evolution of depression disturbances with positive cubic nonlinearity. For elevation disturbances, the results are two-fold. When the cubic nonlinearity is negative, we show that the CV-Whitham equation and the Gardner equation are qualitatively similar, differing only by a small phase lag due to differences in the dispersion term. However, with positive cubic nonlinearity, the differences between the solutions become more pronounced, with the CV-Whitham equation producing sharper waves that suggest the onset of wave breaking.

\end{abstract}


\section{Introduction}
The Korteweg-de Vries (KdV)  equation and its variations have been used in the literature to describe many nonlinear wave phenomena such as soliton interactions, trapped waves, stagnation points beneath solitons among others \cite{Zabusky, Joseph, Johnson, Lax,  Flamarion:2023a, Flamarion:2022d, Flamarion:2023, Smyth:2022}. Such phenomena commonly appears in the study of surface and internal ocean waves. Although this model is rather simple it produces realistic results under the correct regime (weakly nonlinear and weakly dispersive). In internal waves, the nonlinearity is stronger due to features of the density stratification and a similar equation to the KdV with cubic nonlinearity known as the Gardner equation arises a model \cite{Grimshaw:1999, Grimshaw:2010,  Pelinovskii:1998, Shurgalina:2018, Slyunyaev:2001}
\begin{equation}\label{eKdV}
u_{t}+uu_x+\beta u^{2}u_x+ u_{xxx}=0,
\end{equation}
where, $u(x,t)$ is the dimensionless height of the internal wave positioned at $x$ and time $t$. The coefficient $\beta=\pm1$ depends on the stratification. Although both the KdV and Gardner equations are integrable, the Gardner equation describes a more diverse class of solutions. While the KdV equation is known to admit soliton and cnoidal solutions, the Gardner equation also admits breather solutions—wave packets that travel with constant speed and are periodic in both space and time \cite{Chow:2005, Grimshaw:1999, Grimshaw:2010}. 

In a recent study Kharif and Abid \cite{Kharif:2018} and Kharif et al. \cite{Kharif:2017} proposed a heuristic model that became known as the Cubic Vortical Whitham (CV-Whitham) equation which is an asymptotic approximation (in dispersion) to the Gardner equation to investigate nonlinear water waves propagating on a vertically sheared current of constant vorticity in shallow water that satisfies the unidirectional  linear dispersion relation. This model can be written as
\begin{equation}\label{Whitham}
u_{t}-6u_x+uu_x+\beta u^{2}u_x+ {\mathcal{K}}* u_{x}=0.
\end{equation}

The nonlocal operator $K$ is given in terms of the Fourier transform ($\widehat{\mathcal{K}}$)
\begin{equation*}
\widehat{\mathcal{K}}(k)=6\sqrt{\frac{\tanh k}{k}}.
\end{equation*}
This equation extends the classical Whitham equation \cite{Whitham1, Whitham2}, originally introduced as an alternative to the KdV equation for studying wave-breaking, peaking, and short waves. Since its introduction, numerous studies have explored its theoretical aspects. For those interested in further details, we refer to the articles \cite{Smyth:2023, Deconinck, Erm1, Erm2, Flamarion:2022, Flamarion:2022b, Flamarion:2022c, Hur, Hur2, Linares, Moldabayev, Trillo}.

The CV-Whitham equation (\ref{Whitham}) is still weakly nonlinear, but it is fully dispersive since it satisfies the unidirectional linear dispersion relation. Therefore, it represents more nonlinear phenomena such as short waves, peaking and wave breaking, phenomena that the Gardner equation cannot represent. However, the CV-Whitham is more complicated because of its non-integrability. Periodic and solitary waves were computed in the work of Carter et al. \cite{Carter} as well as their stability. For the cubic Whitham equation (neglecting the quadractic nonlinear term) Kalisch \cite{Kalisch:2022} obtained strongly numerical evidences that breather solutions can exist in the cubic Whitham equation.  Solitary wave overtaking collisions were discussed in the work of Flamarion and Pelinovsky \cite{Flamarion:2024}. In this work, the authors investigated  solitary waves collisions to the cubic Whitham equation and showed that it preserves the Lax-geometric categorization \cite{Lax}. Bidirectional-Whitham systems have also appeared in the literature in the context of surface waves and internal waves and have been used accurately to study dispersive wave shocks \cite{Carter0, Kalisch:2019, Vargas:2021, Wang:2022}.

The aim of this work is to study the evolution of initial disturbances through the CV-Whitham equation. We demonstrate that, for negative cubic nonlinearity, the Gardner equation and the CV-Whitham equation yield almost indistinguishable results. However, for positive cubic nonlinearity, the results can be quite different. For instance, a positive disturbance of large amplitude leads to the onset of wave breaking, a phenomenon that the Gardner equation does not capture. An interesting case occurs when the disturbance is negative. In this scenario, there are parameter settings that allow the formation of breather-like structures within the CV-Whitham equation, but not within the Gardner equation.

The structure of the article is as follows: Section 2 outlines the numerical methods. Section 3 presents the results for negative cubic nonlinearity, while Section 4 covers the results for positive cubic nonlinearity. The breather formation results are discussed in Section 5, followed by the conclusions in Section 6.

\section{Numerical Methods}
Solutions to equations (\ref{Whitham}) and (\ref{eKdV}) are computed numerically using the pseudospectral method detailed in \cite{Trefethen}. The computation domain $[-L_x,L_x]$ is chosen with $L_x$ sufficiently large so that the effects of the spatial periodicity, such as the return of small-amplitude radiation is mitigated. Spatial derivatives and the nonlocal operator $\mathcal{K}$ are computed spectrally. For the time advance we use the explicit Runge-Kutta of fourth order with time step $\Delta t$. 

For the initial condition, we consider two types of disturbances a Gaussian pulse
\begin{equation}\label{Gauss}
G(x)= A\exp(-x^2/\sigma^2), 
\end{equation}
where $A$ is the height of initial disturbance and $\sigma$ its width and quasi-rectangular box with smooth slopes
\begin{equation}\label{Box}
B(x)= \frac{A}{2}\Big[\tanh(x+x_0+\gamma)-\tanh(x-x_0+\gamma)\Big].
\end{equation}
where $x_0$ and $\gamma$ are positive constants and $A$ determines the amplitude of the box, which can have positive or negative sign.

\section{Results with negative cubic nonlinearity}

\subsection{Evolution of a Gaussian pulse }

{ 
We recall the analytical results derived from the integrability of the Gardner equation \cite{Pelinovskii:1998, Grimshaw:2010}. If the initial disturbance presents a pure negative pulse (opposite in sign to the quadratic nonlinearity), it evolves into a dispersive wave packet similar to the Airy solution, though nonlinearly deformed. When the initial disturbance is positive but has a small amplitude, it evolves into a set of solitons and a dispersive tail, resembling the KdV processes. If the initial disturbance has a large amplitude (greater than the critical value ($A=1$), the leading soliton assumes a table-top shape. The amplitudes of the generated solitons can be found analytically using the inverse-scattering method \cite{Grimshaw:2002, Osborn:2010}.}

 \begin{figure}[h!]
	\centering	
	\includegraphics[scale =1]{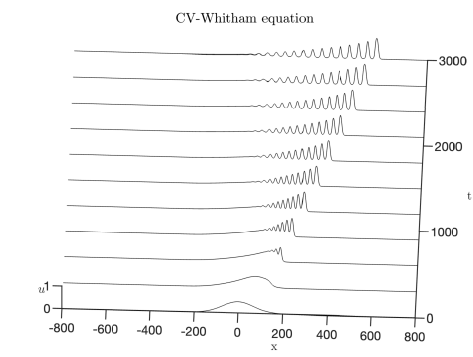}
	\includegraphics[scale =1]{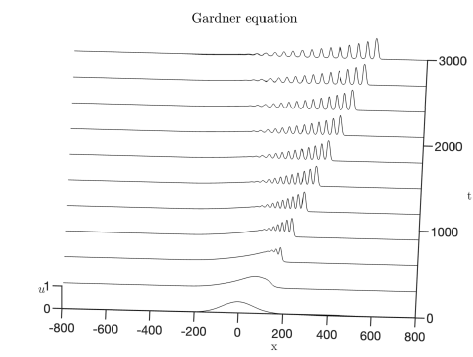}
\caption{Evolution of a Gaussian pulse through the CV-Whitham and Gardner equations. Here, $A=0.5$ and $\sigma=100$ with negative cubic nonlinearity.}
	\label{exp1}
\end{figure}

 \begin{figure}[h!]
	\centering	
	\includegraphics[scale =1]{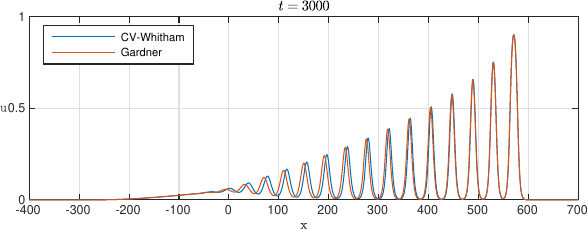}
\caption{Comparison of the evolution of a Gaussian pulse through the CV-Whitham and Gardner equations at time $t=3000$. Here, $A=0.5$ and $\sigma=100$ with negative cubic nonlinearity.}
	\label{exp1comp}
\end{figure}

Figure \ref{exp1} shows the evolution of a wide Gaussian pulse  with small amplitude within the framework of the CV-Whitham equation and the Gardner equation. As we can see the evolution through both equations are apparently the same. In both solutions, a positive initial disturbance decays into a group of solitons  and an oscillating tail. To look closer at their differences we plot in Figure \ref{exp1comp} the solutions at a large time. It shows that not only the solutions are similar qualitatively, but also quantitatively. The only difference between the solutions is the slighly phase lag that Gardner solution has. It is explained by the different dispersion relation that these two equations have. 
 \begin{figure}[h!]
	\centering	
	\includegraphics[scale =1]{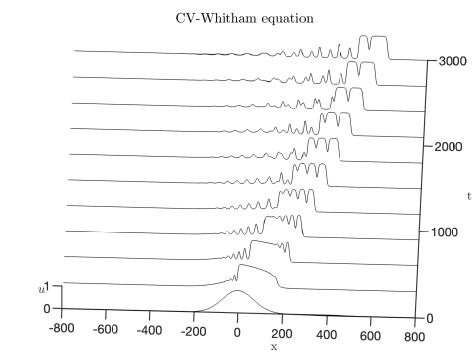}
	\includegraphics[scale =1]{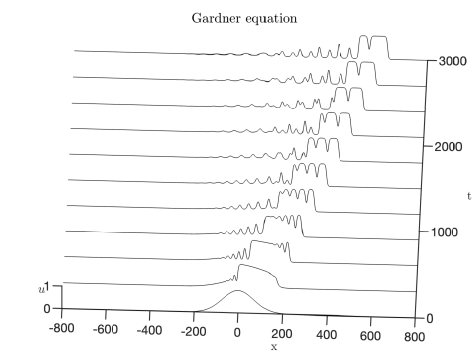}
\caption{Evolution of a Gaussian pulse through the CV-Whitham and Gardner equations. Here, $A=1.0$ and $\sigma=100$ with negative cubic nonlinearity.}
	\label{exp2}
\end{figure}

 \begin{figure}[h!]
	\centering	
	\includegraphics[scale =1]{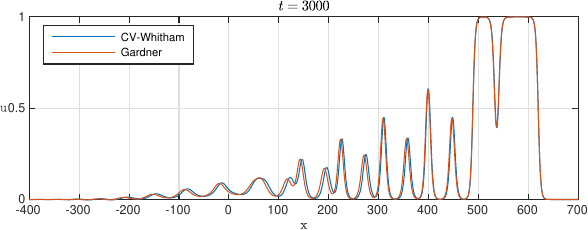}
\caption{Comparison of the evolution of a Gaussian pulse through the CV-Whitham and Gardner equations at time $t=3000$. Here, $A=1$ and $\sigma=100$ with negative cubic nonlinearity.}
	\label{exp2comp}
\end{figure}

\begin{figure}[h!]
	\centering	
	\includegraphics[scale =1]{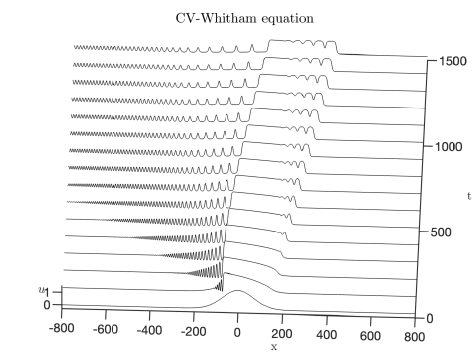}
	\includegraphics[scale =1]{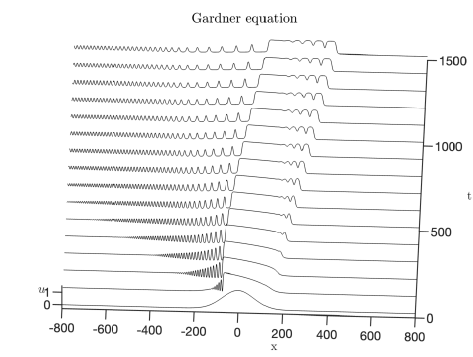}
\caption{Evolution of a Gaussian pulse through the CV-Whitham and Gardner equations. Here, $A=1.5$ and $\sigma=100$ with negative cubic nonlinearity.}
	\label{exp3}
\end{figure}

\begin{figure}[h!]
	\centering	
	\includegraphics[scale =1]{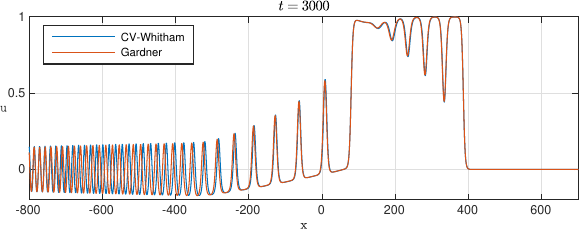}
\caption{Comparison of the evolution of a Gaussian pulse through the CV-Whitham and Gardner equations at time $t=3000$. Here, $A=1$ and $\sigma=100$ with negative cubic nonlinearity.}
	\label{exp3comp}
\end{figure}

As we increase the amplitude of the Gaussian pulse to the limiting soliton solution of the Gardner equation ($A=1$), both solutions change qualitatively, however they still predict the same dynamics. In Figure \ref{exp2}, two sharp declines, or quasi-shocks, occur at the initial stage. This forms a two-step structure resembling a dispersive shock. Solitons with varying polarities emerge, with negative solitons appearing on the higher pedestal and positive ones on the lower pedestal. Subsequently, the negative solitons at the crest of the wave interact with an anti-kink, descending from the limiting soliton, reversing polarity, and merging with the group of solitons that form alongside the anti-kink. In the asymptotic limit, a single limiting soliton and a group of small-scale waves are generated. The expansion of an initial disturbance with an amplitude exceeding the limiting value leads to the formation of a broader, thicker limiting soliton. These thick solitons never exceed the limiting  amplitude given by the Gardner equation.  A comparison between the CV-Whitham equation and the Gardner equation is given in Figure \ref{exp2comp}. As we increase the amplitude of the disturbance similar results are obtained, see Figures \ref{exp3} and \ref{exp3comp}.

\subsection{Evolution of a quasi-rectangular box }
\begin{figure}[h!]
	\centering	
	\includegraphics[scale =1]{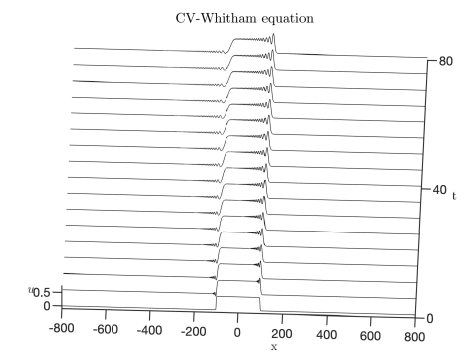}
	\includegraphics[scale =1]{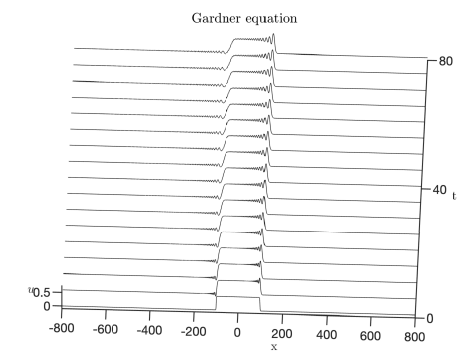}
\caption{Evolution of a box through the CV-Whitham and Gardner equations. Here, $A=0.5$,   $x_0=100$ and $\gamma=0$ with negative cubic nonlinearity.}
	\label{box1}
\end{figure}

\begin{figure}[h!]
	\centering	
	\includegraphics[scale =1]{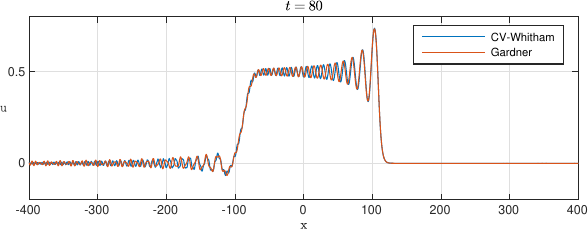}
\caption{Comparison of the evolution of a box through the CV-Whitham and Gardner equations at time $t=80$. Here, $A=0.5$, $x_0=100$ and $\gamma=0$  with negative cubic nonlinearity.}
	\label{box1comp}
\end{figure}

\begin{figure}[h!]
	\centering	
	\includegraphics[scale =1]{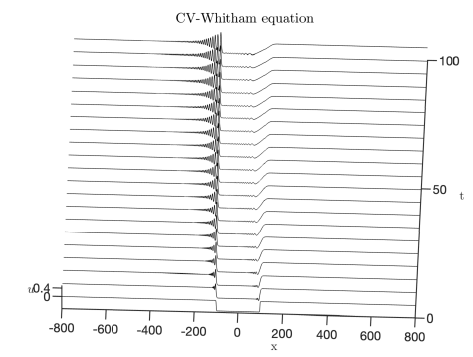}
	\includegraphics[scale =1]{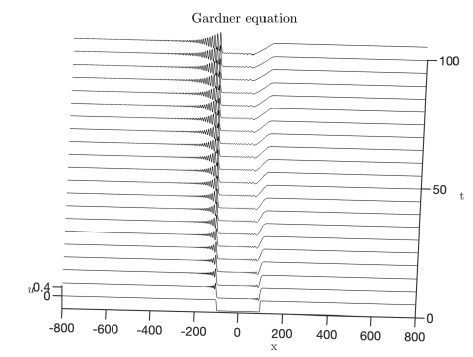}
\caption{Evolution of a box through the CV-Whitham and Gardner equations. Here, $A=-0.5$, $x_0=100$ and $\gamma=0$  with negative cubic nonlinearity.}
	\label{box1neg}
\end{figure}

\begin{figure}[h!]
	\centering	
	\includegraphics[scale =1]{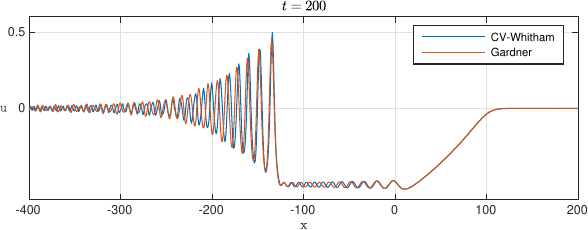}
\caption{Comparison of the evolution of a box through the CV-Whitham and Gardner equations at time $t=200$. Here, $A=-0.5$ and $x_0=100$ with negative cubic nonlinearity.}
	\label{box1compneg}
\end{figure}

In this case, the wave dynamics is rather simple compared to the evolution of Gaussian pulses. Therefore, we discuss this case briefly. For a positive box, a leading solitary wave appears as a wave front accompanied by a dispersive tails formed on the top of the box. { The amplitudes of the generated solitons can be found analytically in the frame of Gardner equation \cite{Grimshaw:2002}. }These waves move to the zero mean level in the form of short waves, see for instance Figures \ref{box1} and \ref{box1comp}. Meanwhile, the evolution of a negative box is featured by the generation of a dispersive wave train and a hole where the initial disturbance is positioned. Even so, the CV-Whitham and the Gardner solutions agree well as shown in Figures  \ref{box1neg} and \ref{box1compneg} expected by a phase lag.

The main difference occurs due to the generation of short waves. Such waves are not well captured in the Gardner equation (long-wave model) and such waves travel faster than the CV-Whitham short waves.

\newpage
\newpage
\newpage

\section{Results with positive cubic nonlinearity}

\subsection{Evolution of positive disturbances }
{ The Gardner equation with positive cubic nonlinearity is also solved by the inverse-scattering method, but its solution is more complicated, see for instance \cite{Osborn:2010}. For the positive initial disturbances its evolution leads to the generation of the solitons (small amplitude solitons as KdV solitons, and large-amplitude solitons as mKdV solitons). However, if the initial disturbance is negative or sign-variable, its evolution leads to the generation of solitons and breathers.}

\begin{figure}[h!]
	\centering	
	\includegraphics[scale =1]{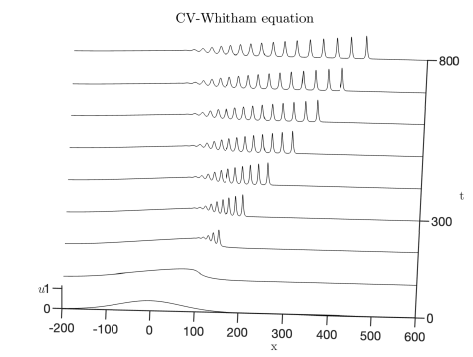}
	\includegraphics[scale =1]{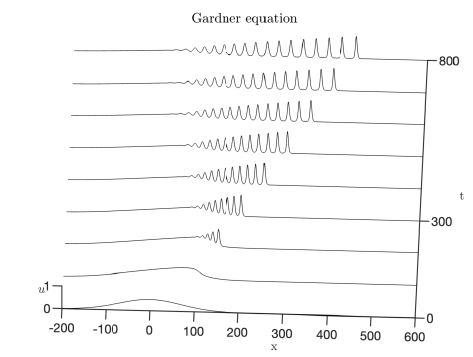}
\caption{Evolution of a Gaussian pulse through the CV-Whitham and Gardner equations. Here, $A=0.5$ and $\sigma=100$ with positive cubic nonlinearity.}
	\label{positiveexp1}
\end{figure}

\begin{figure}[h!]
	\centering	
	\includegraphics[scale =1]{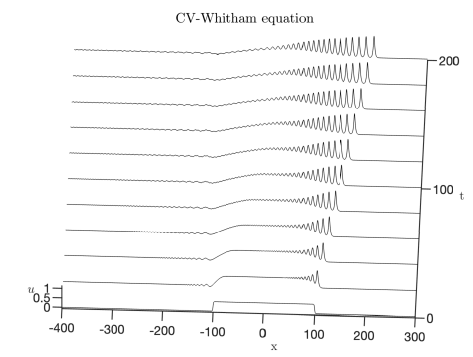}
	\includegraphics[scale =1]{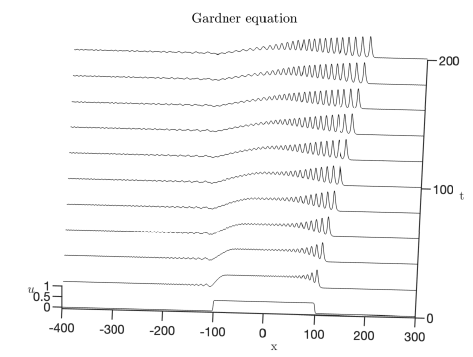}
\caption{Evolution of a Gaussian pulse through the CV-Whitham and Gardner equations. Here, $A=0.5$ and $x_0=100$ with positive cubic nonlinearity.}
	\label{positivebox2}
\end{figure}

\begin{figure}[h!]
	\centering	
	\includegraphics[scale =1]{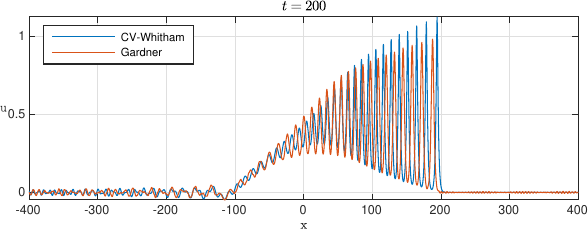}
\caption{Comparison of the evolution of a box through the CV-Whitham and Gardner equations at time $t=200$. Here, $A=0.5$ and $x_0=100$ with positive cubic nonlinearity.}
	\label{positivebox1comp}
\end{figure}

\begin{figure}[h!]
	\centering	
	\includegraphics[scale =1]{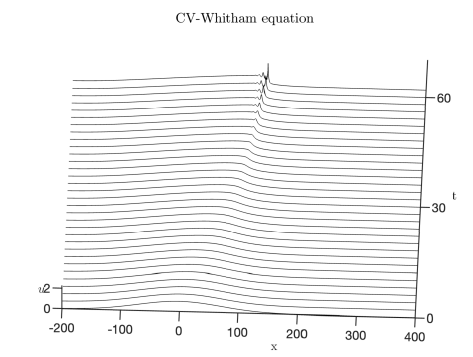}
	\includegraphics[scale =1]{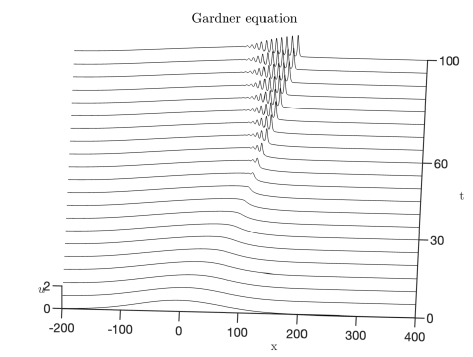}
\caption{Evolution of a Gaussian pulse through the CV-Whitham and Gardner equations. Here, $A=1.0$ and $\sigma=100$ with positive cubic nonlinearity.}
	\label{positiveexp2}
\end{figure}

The results on evolution of positive disturbances is straightforward. For small amplitude disturbances (Gaussian pulses or boxes) the CV-Whitham equation and the Gardner equation yield qualitatively the same solution except for a phase lag in the Gardner equation and slightly more pronounced crests on the CV solitons. Comparisons of both models for Gaussian pulses and smooth boxes are shown in Figures \ref{positiveexp1}-\ref{positivebox1comp}. The main differences arise when we increase the disturbance amplitude. In this case, short waves are generated and an onset of wave breaking takes place. Figures \ref{positiveexp2} and \ref{positiveexp2comp} display the evolution Guassian pulse of amplitude $A=1$, while the Gardner equation predicts the evolution of a wave train of solitons the leading wave in the CV-Whitham equation becomes sharp leading to an onset of wave breaking. This occurs because differently from the Gardner equation the CV-Whitham equation captures intermediate and short waves.

 \begin{figure}[h!]
	\centering	
	\includegraphics[scale =1]{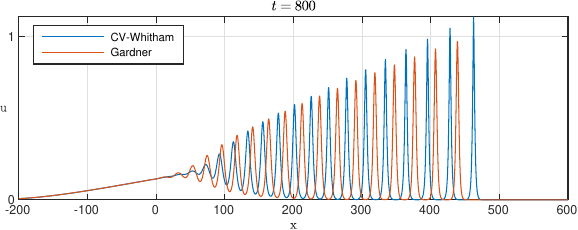}
\caption{Comparison of the evolution of a Gaussian pulse through the CV-Whitham and Gardner equations at time $t=3000$. Here, $A=0.5$ and $\sigma=100$ with positive cubic nonlinearity.}
	\label{positiveexp2comp}
\end{figure}

 \begin{figure}[h!]
	\centering	
	\includegraphics[scale =1]{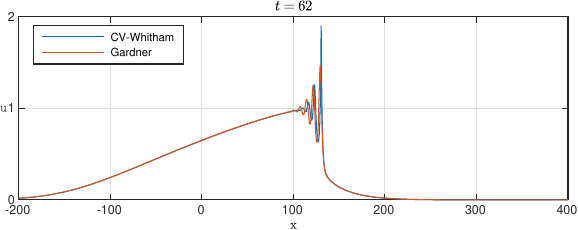}
\caption{Comparison of the evolution of a Gaussian pulse through the CV-Whitham and Gardner equations at time $t=62$. Here, $A=1.0$ and $\sigma=100$ with positive cubic nonlinearity.}
	\label{positiveexp2comp}
\end{figure}

\section{Breather generation}
\begin{figure}[h!]
	\centering	
	\includegraphics[scale =1]{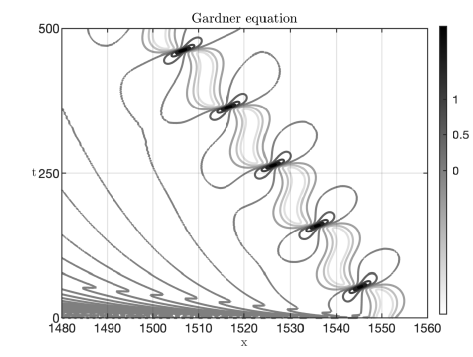}
	\includegraphics[scale =1]{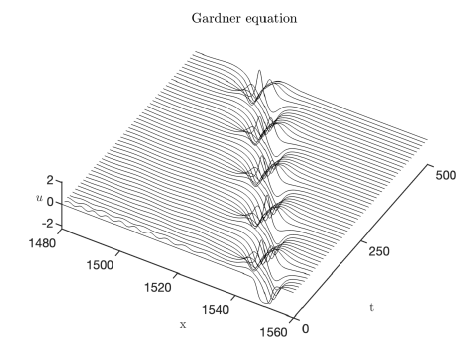}
\caption{Breather formation in the Gardner equation. The evolution of a negative box. Here, $A=-2$, $x_0=\sqrt{6}$ and $\gamma=1550$ with positive cubic nonlinearity.}
	\label{Grimshaw}
\end{figure}

In this section, we focus on breather-like structures. Grimshaw et al. \cite{Grimshaw:2010} investigated the evolution of box-like disturbances within the Gardner equation. They found that when the initial disturbance has the same polarity as the quadratic nonlinear coefficient, only solitons are generated. However, if the initial disturbance has the opposite polarity, a variety of outcomes can occur, including solitons of different polarities and breathers. Theoretical conditions for the appearance of breather-like structures were determined based on the amplitude and width of the initial disturbance. { More precisely, the authors considered the Gardner equation scaled as 
\begin{equation}\label{GardnerEfim}
u_{t}+6uu_x+6u^{2}u_x+ u_{xxx}=0
\end{equation}
and the initial disturbance defined as
\begin{equation}\label{boxd}
u(x,0)=\begin{cases}
             A_0  & \text{if } |x| < L/2, \\
             0  & \text{if } |x| \ge L/2.
       \end{cases} \quad
\end{equation}
They showed through the inverse scattering problem that negative disturbances can give birth to both solitons and breathers. In particular for $A_0=-2$ a breather should appear for $0.9<L<2.0$. Considering our scale (the coefficients of the quadratic and cubic nonlinearity are equal to 1), it means that breathers give birth
then the disturbance evolves into a breather structure if $A_0=-2$  and $0.9\sqrt{6}=2.20<L<4.89=2\sqrt{6}$.} In this work, the box (\ref{boxd}) is approximated by the quasi-rectangular box (\ref{Box}). Figure \ref{Grimshaw} shows the evolution of a smooth negative box within the Gardner framework. As observed, the box evolves into a breather-type structure with small dispersion in the tail.

In the previous sections, we demonstrated that the CV-Whitham and Gardner equations may not produce qualitatively identical results, primarily because the Gardner equation neglects low frequencies. Here, we observe a similar phenomenon. When we use the same disturbance as the initial condition for the CV-Whitham equation, the solution breaks in a short time, and no breather-like structure forms. Therefore, we explore different parameter regimes where breather-like structures may arise.
 \begin{figure}[h!]
	\centering	
	\includegraphics[scale =1]{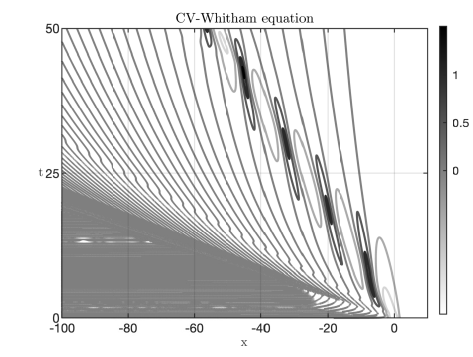}
	\includegraphics[scale =1]{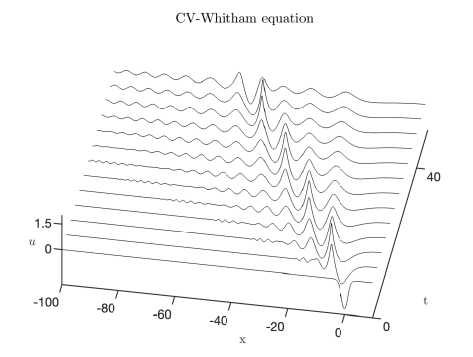}
\caption{Breather formation in the CV-Whitham. The evolution of a negative box. Here, $A=-2$, $x_0=1.2$  and $\gamma=0$ with positive cubic nonlinearity.}
	\label{breathercv1}
\end{figure}

\begin{figure}[h!]
	\centering	
	\includegraphics[scale =1]{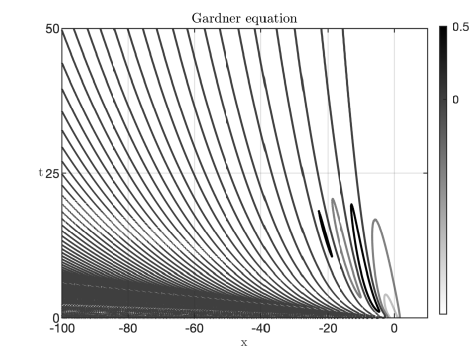}
	\includegraphics[scale =1]{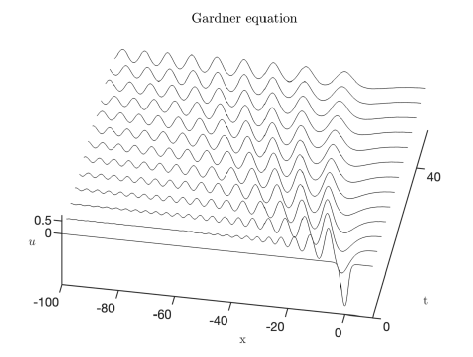}
\caption{The evolution of a negative box within the Gardner equation. Here, $A=2$, $x_0=1.2$ and $\gamma=0$ with positive cubic nonlinearity.}
	\label{breatherge1}
\end{figure}

By varying the set of parameters ($A$ and $x_0$) we can identify breather-like structures for the CV-Whitham equation. Figure \ref{breatherge1} illustrates a breather-like structure that emerges within the CV-Whitham equation. Note that while the disturbance evolves into a simple breather-like structure with a dispersive tail, the same disturbance in the Gardner equation results in a dispersive wave train. 

\begin{figure}[h!]
	\centering	
	\includegraphics[scale =1]{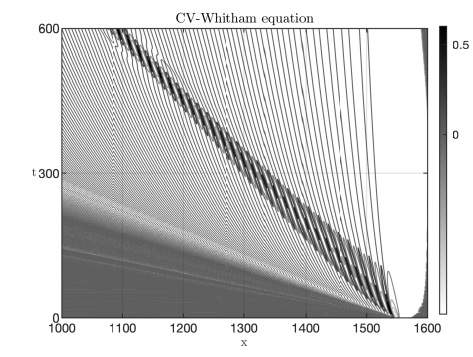}
	\includegraphics[scale =1]{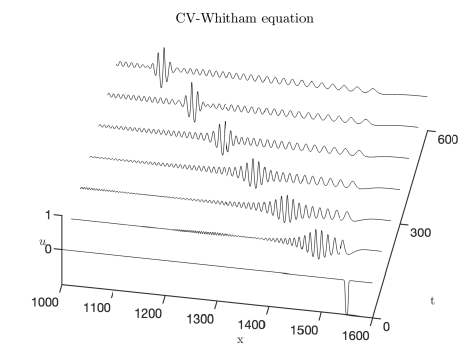}
\caption{Breather formation in the CV-Whitham. The evolution of a negative box. Here, $A=1$, $x_0=3$ and $\gamma=1550$ with positive cubic nonlinearity.}
	\label{breathercv0}
\end{figure}

\begin{figure}[h!]
	\centering	
	\includegraphics[scale =1.1]{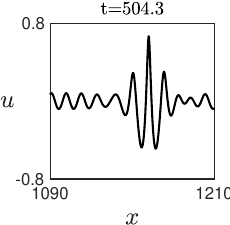}
	\includegraphics[scale =1.1]{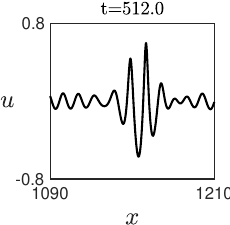}
	\includegraphics[scale =1.1]{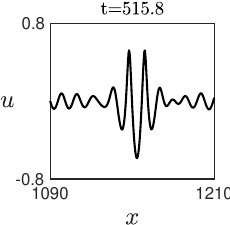}
	\includegraphics[scale =1.1]{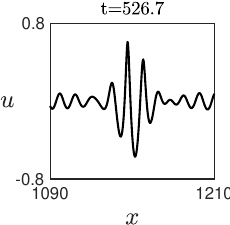}
	\includegraphics[scale =1.1]{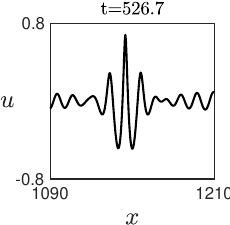}
	\includegraphics[scale =1.1]{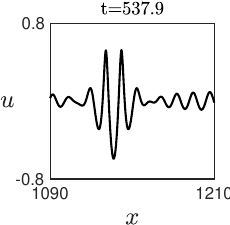}
\caption{Snapshots of the Breather formation in the CV-Whitham described in Figure \ref{breathercv0}. }
	\label{snap1}
\end{figure}

\begin{figure}[h!]
	\centering	
	\includegraphics[scale =1]{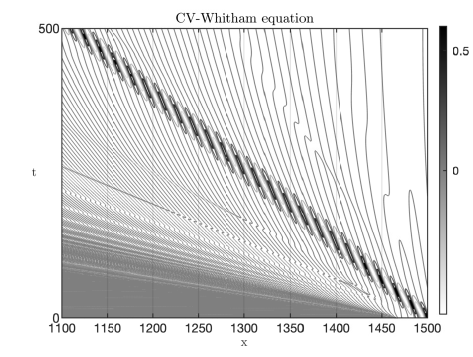}
	\includegraphics[scale =1]{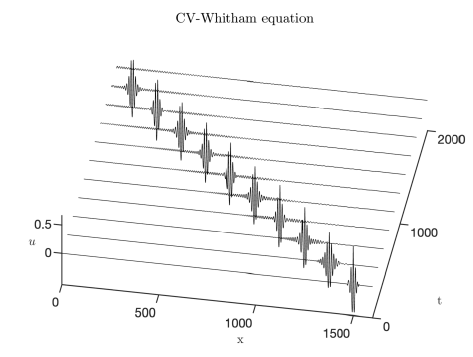}
\caption{Breather evolution in the CV-Whitham. 
At time $ = 504.3$, the breather-like structure shown in Figure \ref{snap1} is initialized with its maximum at  $x = 1500$, and its dispersive tail is zeroed by multiplying by a smooth function.}
	\label{breathercv00}
\end{figure}

A more complex breather structure arises when we decrease \(A\) to 1 and use a broader box with \(x_0 = 3\). Figure \ref{breathercv0} shows the evolution of this smooth negative box. Initially, the box evolves into a dispersive wave train. At intermediate times, a well-formed breather structure emerges in the wave field, and at later times, it detaches from the wave train to form a distinct breather structure. Snapshots at different times are depicted in Figure \ref{snap1}. To further examine this structure, we reset the wave field at time \(t = 504\) with its peak at \(x = 0\), manually zeroing the dispersive tail, and evolve it again using the CV-Whitham equation. The results, shown in Figure \ref{breathercv00}, reveal that the initial data evolves into a well-defined breather structure. To the best of our knowledge, this is the first report of breather structures within the CV-Whitham equation in the literature.

\section{Conclusions}
In this work, we studied the evolution of disturbances within the CV-Whitham equation and compared the results with the Gardner equation. We showed that when the CV-Whitham equation has a negative cubic nonlinear term, the evolution is qualitatively similar to the Gardner equation. However, with a positive cubic nonlinearity, two scenarios emerge: (i) for positive disturbances, the main difference is a phase lag, but as we increase the amplitude, solutions to the CV-Whitham equation exhibit the onset of wave breaking; (ii) for negative disturbances, the solutions can differ qualitatively. For instance, breather-like structures may appear in one model but not in the other.

\section{Acknowledgements}
{The work of E.P was supported by RSF 24-47-02007 (section 4 and 5).}

%

	\section*{Declarations}
	
	\subsection*{Conflict of interest}
	The authors state that there is no conflict of interest. 
	\subsection*{Data availability}
	
	Data sharing is not applicable to this article as all parameters used in the numerical experiments are informed in this paper.

\end{document}